\documentclass[10pt,conference]{IEEEtran}
\IEEEoverridecommandlockouts
\usepackage{cite}
\usepackage{amsmath,amssymb,amsfonts}
\usepackage{algorithmic}
\usepackage{graphicx}
\usepackage{textcomp}
\usepackage{xcolor}
\usepackage{multirow}
\usepackage{soul}
\usepackage{verbatim} 
\usepackage[font=small,skip=2pt]{caption}
\def\BibTeX{{\rm B\kern-.05em{\sc i\kern-.025em b}\kern-.08em
5    T\kern-.1667em\lower.7ex\hbox{E}\kern-.125emX}}

\begin{document}

\title{MPNA: A Massively-Parallel Neural Array Accelerator with Dataflow Optimization for Convolutional Neural Networks\\
}

%\author{\IEEEauthorblockN{Muhammad Abdullah Hanif}
%\IEEEauthorblockA{\textit{Institute of Computer Engineering} \\
%\textit{Vienna University of Technology (TU Wien))}\\
%Vienna, Austria \\
%muhammad.hanif@tuwien.ac.at}
%\and
%\IEEEauthorblockN{Rachmad Vidya Wicaksana Putra}
\author{\IEEEauthorblockN{Muhammad Abdullah Hanif,\textsuperscript{1,*} Rachmad Vidya Wicaksana Putra,\textsuperscript{1,*} Muhammad Tanvir,\textsuperscript{2} Rehan Hafiz,\textsuperscript{2} \\
Semeen Rehman,\textsuperscript{1} Muhammad Shafique\textsuperscript{1}}
\IEEEauthorblockA{\textit{\textsuperscript{1}Vienna University of Technology (TU Wien)}, Vienna, Austria \\ 
\textit{\textsuperscript{2}Information Technology University (ITU)}, Lahore, Pakistan \\
(muhammad.hanif, rachmad.putra, semeen.rehman, muhammad.shafique)@tuwien.ac.at, (msee17004, rehan.hafiz)@itu.edu.pk} 
\thanks{*The authors contributed equally to this work}}

\maketitle

\begin{abstract}
The state-of-the-art accelerators for Convolutional Neural Networks (CNNs) typically focus on accelerating {\it only} the convolutional layers, but do not prioritize the fully-connected layers much. Hence, they lack a {\it synergistic} optimization of the hardware architecture and diverse dataflows for the complete CNN design, which can provide a higher potential for performance/energy efficiency.
Towards this, we propose a novel {\it Massively-Parallel Neural Array (MPNA)} accelerator that integrates two heterogeneous systolic arrays and respective highly-optimized dataflow patterns to {\it jointly} accelerate both the convolutional (CONV) and the fully-connected (FC) layers. Besides fully-exploiting the available off-chip memory bandwidth, these optimized dataflows enable high data-reuse of all the data types (i.e., weights, input and output activations), and thereby enable our MPNA to achieve high energy savings.
We synthesized our MPNA architecture using the ASIC design flow for a $28$ nm technology, and performed functional and timing validation using multiple real-world complex CNNs. MPNA achieves $149.7$ GOPS/W at $280$ MHz and consumes $239$ mW. Experimental results show that our MPNA architecture provides $1.7\times$ overall performance improvement compared to state-of-the-art accelerator, and $51$\% energy saving compared to the baseline architecture.
\end{abstract}

%%%%%%%%%%%%%%%%%%%%%%%%%%%%%%%%%%%%%%%%%%%%%%%%%%%%%%%%%%%%%%%%%%%%%%%%%%%%%%%%%%%%%%%%%%%%%%%%%%%%%%%%%%%%%%
\section{Introduction}
\label{sec:Introduction}

Machine learning has rapidly proliferated into different field of life ranging from automotive and smart environments to medicine. 
Due to their high accuracy, larger and deeper Convolutional Neural Networks (CNNs) have become the key technology for many applications like advanced vision processing. However, it comes at the cost of significant computational and energy requirements, and thus, requiring  specialized accelerator-based architectures. 

%%%%%%%%%%%

\subsection{State-of-the-Art and Their Limitations} 
Plenty of work has been carried out in designing CNN accelerators~\cite{Sze_DNNsurvey_IEEE17}, which mostly focus on~\textbf{un-structurally sparse neural networks} like {\it EIE}~\cite{Han_EIE_ISCA16}, {\it SCNN}~\cite{Parashar_SCNN_ISCA17} and {\it Cnvlutin}~\cite{Albericio_Cnvlutin_ISCA16}. 
The {\it EIE} and {\it SCNN} architectures make use of both sparsity in weights and activations for accelerating the networks, while {\it Cnvlutin} exploits sparsity in activations only. 
The sparsity in activations usually comes from converting all the negative activations to zeros in the Rectified Linear Unit (ReLU), as used in these designs. However, the above accelerators cannot efficiently handle advanced activation functions like Leaky-ReLU that do not result in high sparsity to help the training process~\cite{Klambauer_SELU_NIPS17}~\footnote{Other non-linear activation functions like scaled exponential linear units (SELU) are not in the scope of this work.}.
Tensor Processing Unit (TPU)~\cite{Jouppi_GoogleTPU_ISCA17}, DaDianNao~\cite{Luo_DaDianNao_TC17} and Eyeriss~\cite{Chen_Eyeriss_JSSC16} architectures accelerate \textbf{dense neural networks}, and can also be modified to provide support for structurally sparse networks~\cite{Anwar_StructPrune_arXiv15}. 
\textit{Although these architectures show good performance in terms of latency for the convolutional (CONV) layers, they offer very limited acceleration for the fully-connected (FC) layers}, as we will show with the help of a motivational case study in the following.

\begin{figure}[t]
\centering
\includegraphics[width=0.8\linewidth]{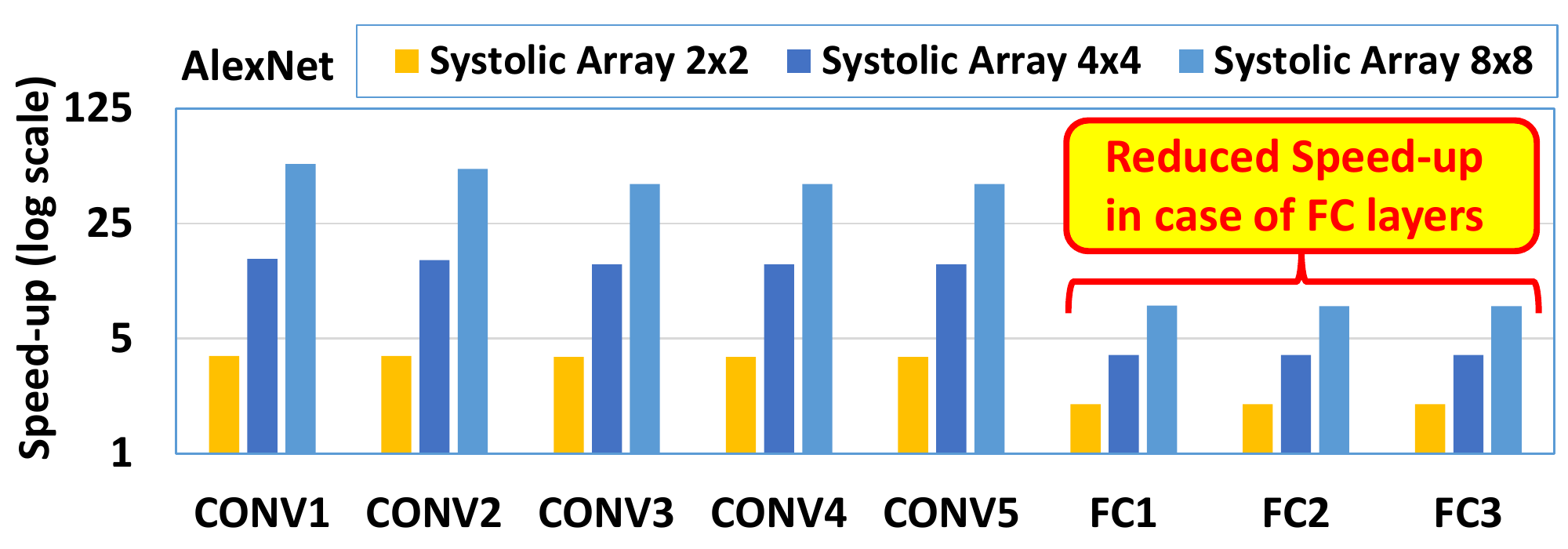}
\caption{Speedup for CONV and FC layers of the AlexNet for different sizes of systolic array normalized to a 1x1 systolic array system.}
\label{fig:Fig_SA_n_Analysis_AlexNet}
\end{figure}

\begin{figure}[t]
\centering
\includegraphics[width=1\linewidth]{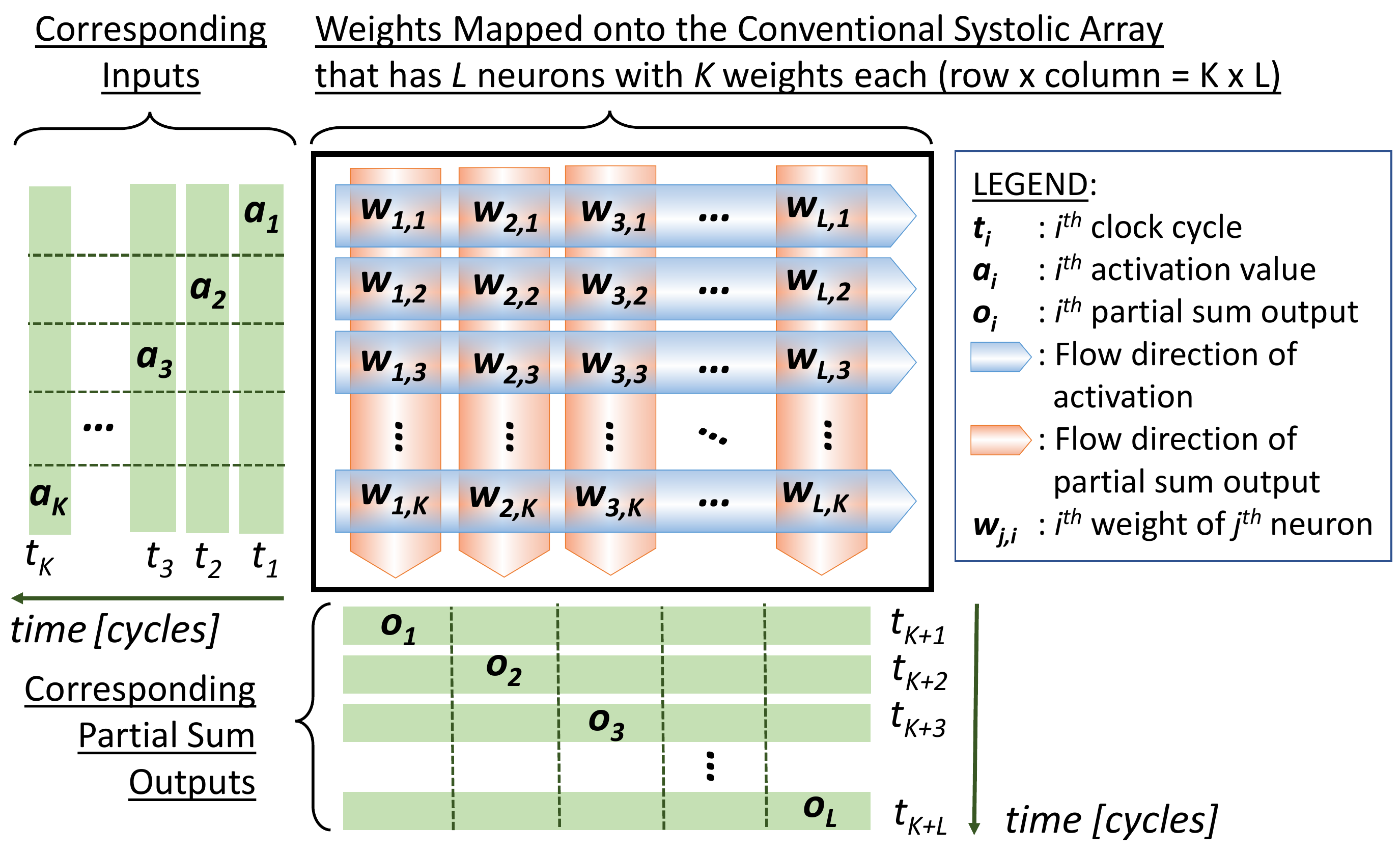}
\caption{Dataflow for the execution of FC layers in conventional SA. The inputs are fed to SA from the left, and shifted one step to the right at each cycle. The computed partial sums are shifted downwards every cycle. The computed partial sums start appearing at the output after $K+1$ cycles.}
\label{fig:Dataflow_ConventionalSA}
\end{figure}

\subsection{Motivational Case-Study} To achieve high performance and power/energy efficiency, state-of-the-art CNN accelerators exploit the reuse of activations, weights and partial sums, thereby increasing the data locality and reducing the number of off-chip memory accesses~\cite{Chen_Eyeriss_JSSC16}. 
In this respect, the conventional systolic array-based designs (like Google's TPU~\cite{Jouppi_GoogleTPU_ISCA17}) render very effective, because each Processing Element (PE) in the Systolic Array (SA) performs three key tasks.
(1) It receives data from their upstream neighbor(s). 
(2) It performs the basic multiply-and-accumulate (MAC) operation(s).
(3) It passes the data along with the partial result(s) to their downstream neighbor(s). 
Hence, for computations that involve both activation and weight reuse (i.e., CONV layer), the overall speedup of these systolic arrays is significant as shown in Fig.~\ref{fig:Fig_SA_n_Analysis_AlexNet} for AlexNet~\cite{Krizhevsky_AlexNet_NIPS12}. 
However, in case of only activation reuse -- i.e., where one single input has to be used for multiple computations while the weights have to be used only once -- the amount of speedup is very limited; see Fig.~\ref{fig:Fig_SA_n_Analysis_AlexNet}. 
Such operations are excessively found in the FC layers, as illustrated by their dataflow in Fig.~\ref{fig:Dataflow_ConventionalSA}.

Our analysis in Fig.~\ref{fig:Fig_SA_n_Analysis_AlexNet}b illustrates that although the overall speedup for the CONV layers is significant, the conventional systolic array does not provide matching speedup for the FC layers. This ultimately limits the overall performance of accelerating the networks, especially when dominated by FC layers. \textit{Hence, there is a significant need for an accelerator-based architecture that can expedite {\bf both} the CONV and the FC layers, to achieve a high speedup for the complete CNN.}  Designing such an architecture, however, bears a broad range of challenges, as discussed below.

%%%%%%%%%%%
\subsection{Associated Scientific Challenges} Firstly, specialized systolic arrays need to be designed that can accelerate both CONV and FC layers without incurring significant area and power/energy overheads compared to the conventional approaches. 
Such systolic arrays should account for diverse dataflow of both types of layers, while fully utilizing the available memory bandwidth. 
For instance, the CONV layers' acceleration needs simple, fast yet massively-parallel PEs to exploit activation, weight, and partial sums reuse. While, the FC layers' acceleration can only exploit activation reuse in a single-sample batch processing. 
Note: the FC layers acceleration can exploit weight reuse only in a multi-sample batch processing, which is not suitable for real-time or latency-sensitive applications. 

%%%%%%%%%%%

\subsection{Our Novel Contributions} 
To overcome the above research challenges, we make the following novel contributions.

\begin{itemize}
\item \textbf{MPNA: A \underline{M}assively-\underline{P}arallel \underline{N}eural \underline{A}rray (Section~\ref{sec:MPNA})}: 
It integrates heterogeneous systolic arrays, an efficient dataflow controller and specialized on-chip memory to maximize data reuse, and other necessary architectural components to jointly accelerate FC and CONV layers. 

\item \textbf{A Design Methodology (Section~\ref{Sec: Methodology}):} 
The MPNA architecture is systematically designed using a synergistic
methodology that explores different data reuse techniques and architectural alternatives. Towards this, we also present the computational complexity and data-reuse analysis for CNNs (Section~\ref{sec:Analysis}).

\item \textbf{Optimized Dataflows (Section~\ref{sec:Dataflow}):} 
We propose different dataflow optimizations for efficient processing on heterogeneous systolic arrays while reducing the number of DRAM accesses and maximally using the data reuse, thereby improving the overall processing efficiency.

\item \textbf{Hardware Implementation and Evaluation (Section~\ref{sec:Evaluation} and \ref{sec:Results}):} 
We synthesize the complete MPNA architecture for a 28nm CMOS technology library using the ASIC design tools, and perform functional and timing validation. 
Our results show that the MPNA architecture offers $1.7\times$ overall performance improvement compared to state-of-the-art accelerator, and $51$\% energy saving compared to the baseline architecture. MPNA achieves $149.7$ GOPS/W performance efficiency at $280$ MHz and consumes $239$ mW.
\end{itemize}

%%%%%%%%%%%%%%%%%%%%%%%%%%%%%%%%%%%%%%%%%%%%%%%%%%%%%%%%%%%%%%%%%%%%%%%%%%%%%%%%%%%%%%%%%%%%%%%%%%%%%%%%%%%%%%

\section{Preliminaries}
\label{Sec: Preliminaries}
Before proceeding further, first, we present basics of CNNs, which are necessary to understand the contributions in the later sections.

Neural networks are composed of various layers, which are connected in cascade. Each layer receives some input from the preceding layer, performs certain operations, and forwards the result to the succeeding layer. A CNN mainly consists of four types of processing layers: (1) {\it Convolutional,} for extracting features; (2) {\it Fully-Connected,} for classification; (3) {\it Activation,} for introducing non-linearity; and (4) {\it Pooling,} for sub-sampling. Among these layers, the convolutional (CONV) layers are the most computationally intensive, while the fully-connected (FC) layers are the most memory intensive ones. Fig.~\ref{fig:Fig_AlexNet_VGG_DataReuse}a illustrates this observation using the percentage of weights and MAC operations required for the CONV and FC layers in the {\it AlexNet} and {\it VGG-16} networks. 

\begin{figure}[t]
\centering
\includegraphics[width=0.90\linewidth]{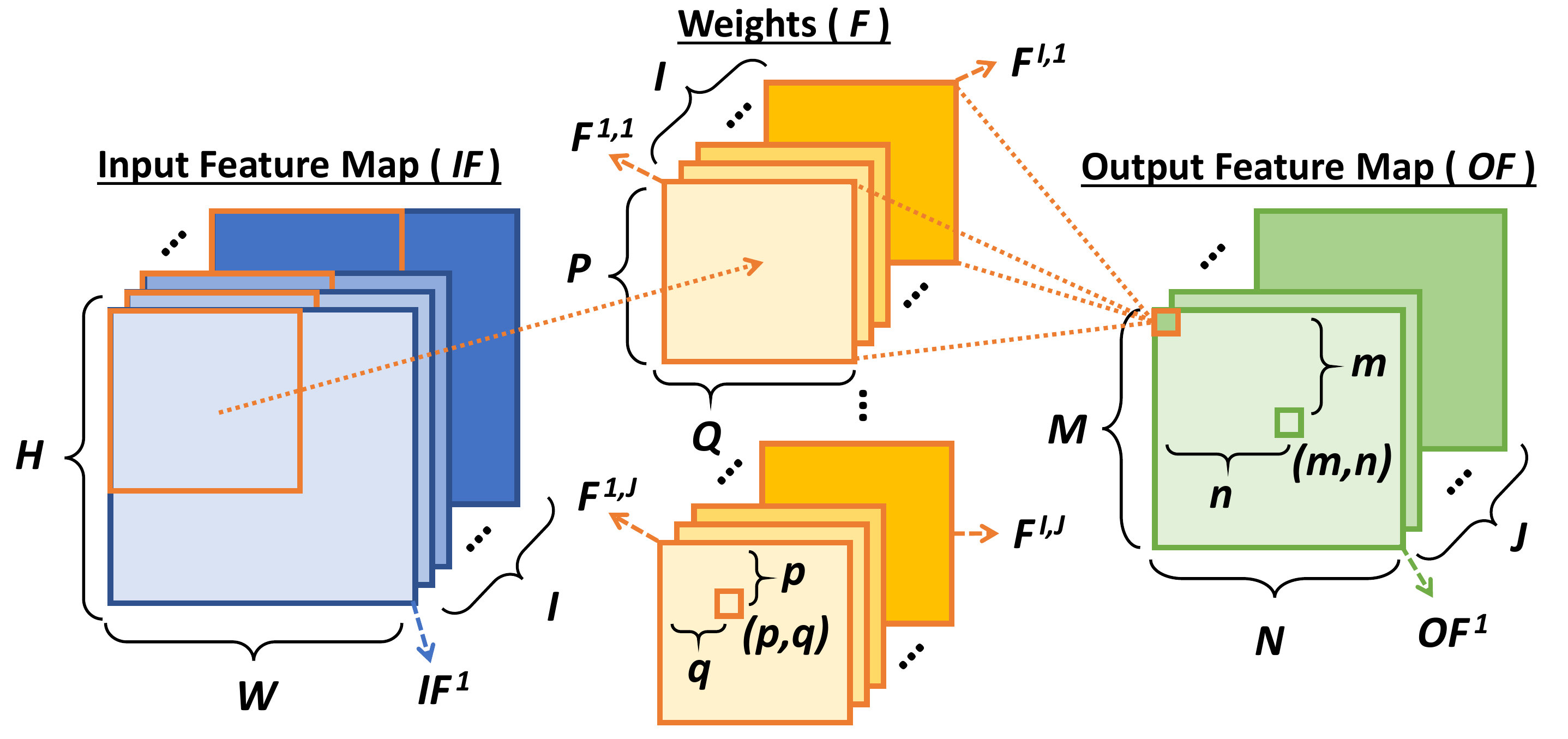}
\caption{Illustration of a single CONV layer, where a set of input feature maps is convolved with the filters to generate the output feature maps.}
\label{fig:Fig_CONV_Illustration}
\end{figure}

A convolutional layer receives the data from the inputs or the preceding layer, and performs the convolution operation using several filters to obtain several output feature maps, each corresponding to the output of one filter. Fig.~\ref{fig:Fig_CONV_Illustration} shows a detailed illustration of a single convolutional layer.
Here, $IF^{i}$ is the $i^{th}$ 2D input feature map, $OF^{j}$ is the $j^{th}$ 2D output feature map, $F^{i,j}$ is the 2D kernel of a filter between $IF^{i}$ and $OF^{j}$. The term $OF^{j}(m,n)$ denotes the activation at location (m,n) of the $j^{th}$ output feature map, i.e., $OF^{j}$. Similarly, $F^{i,j}(p,q)$ denotes the weight/synapse at location (p,q) in the 2D filter kernel between $IF^{i}$ and $OF^{j}$. 
Let's consider convolutional stride = 1, unless stated otherwise. The FC layers can be considered as a special case of CONV layers where the input and output is a 1D array and, therefore, can be represented by the above terminologies.

%%%%%%%%%%%%%%%%%%%%%%%%%%%%%%%%%%%%%%%%%%%%%%%%%%%%%%%%%%%%%%%%%%%%%%%%%%%%%%%%%%%%%%%%%%%%%%%%%%%%%%%%%%%%%%
\section{Methodology for Designing DNN Accelerators}
\label{Sec: Methodology}
\begin{figure}[htbp]
\centering
\includegraphics[width=0.80\linewidth]{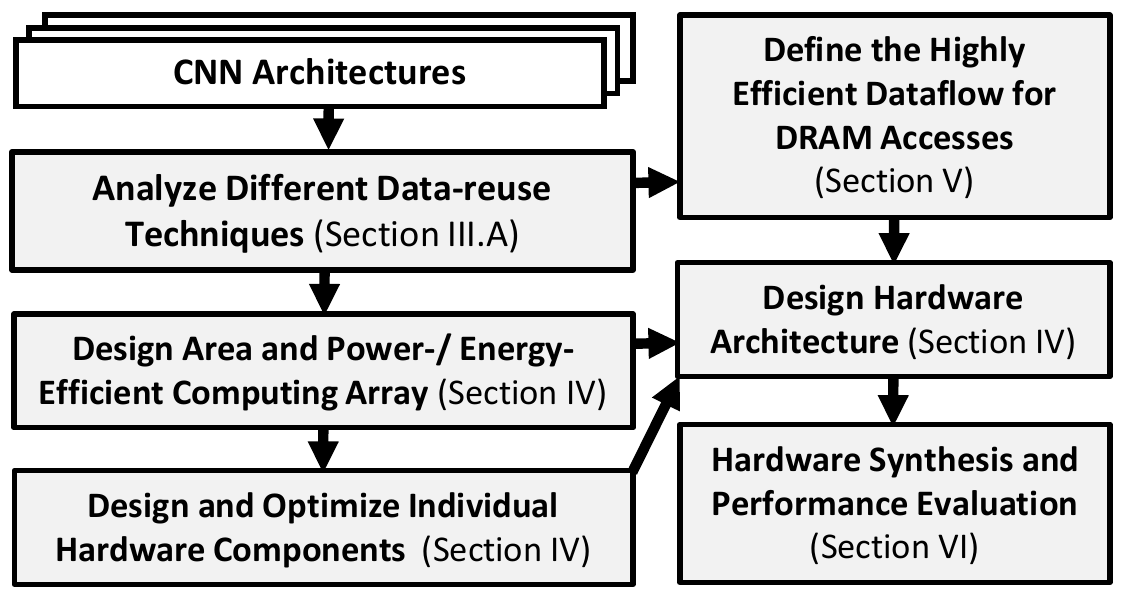}
\caption{Overview of our MPNA design methodology showing key steps.} 
\label{fig:Fig_MPNA_DesignMethod}
\end{figure}

Our methodology for designing optimized DNN accelerator-based architectures is shown in Fig. \ref{fig:Fig_MPNA_DesignMethod}. It consists of the following key steps, which are explained in detail in the subsequent sections.

\begin{enumerate}

\item Analyze different data-reuse techniques, which can be exploited for energy- and performance-efficient execution of a given DNN on the hardware accelerator.
\item Design and optimize individual hardware components for all the elementary functions required for the DNN execution.
\item Design area and power/energy-efficient processing arrays as key accelerator units, which can support the most effective types of dataflow/parallelism for high-performance execution of all the computational layers of a given DNN.
\item Devise an optimized hardware configuration considering key architectural parameters like the size and number of processing arrays, interconnect of components affecting the supported dataflows, on-chip buffers, data reuse and memory organization considering the available DRAM bandwidth.
\item Define highly efficient dataflows for reducing the total number of DRAM accesses required for the DNN inference.
\item Synthesis of the complete hardware architecture for detailed benchmarking for area, performance/throughput,  and power/energy consumption considering different DNN configurations.
\end{enumerate}

\subsection{Computational Complexity and Data-Reuse Analysis}
\label{sec:Analysis}
\begin{figure}[htbp]
\centering
\includegraphics[width=0.9\linewidth]{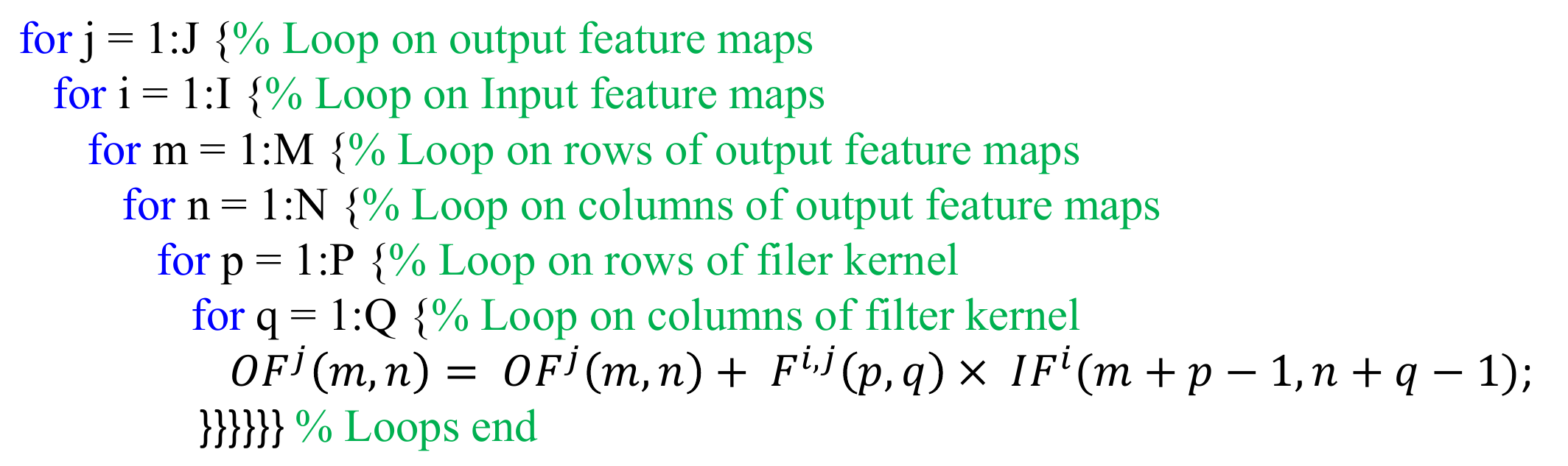}
\caption{Pseudocode for the processing of CONV layers.}
\label{fig:convolution_loops}
\end{figure}

The CNN complexity can be estimated by analyzing the number of computations required for the CONV and FC layers. Fig.~\ref{fig:convolution_loops} illustrates the pseudocode of the CNN layer execution. Here, $I$ and $J$ define the number of input and output feature maps; $M$ and $N$ define the number of rows and columns in the output feature maps; and $P$ and $Q$ define the number of rows and columns in filter kernels. These parameters can be used to define other parameters, like the the number of filters can be derived from $J$.

\begin{figure}[t]
\centering
\includegraphics[width=1\linewidth]{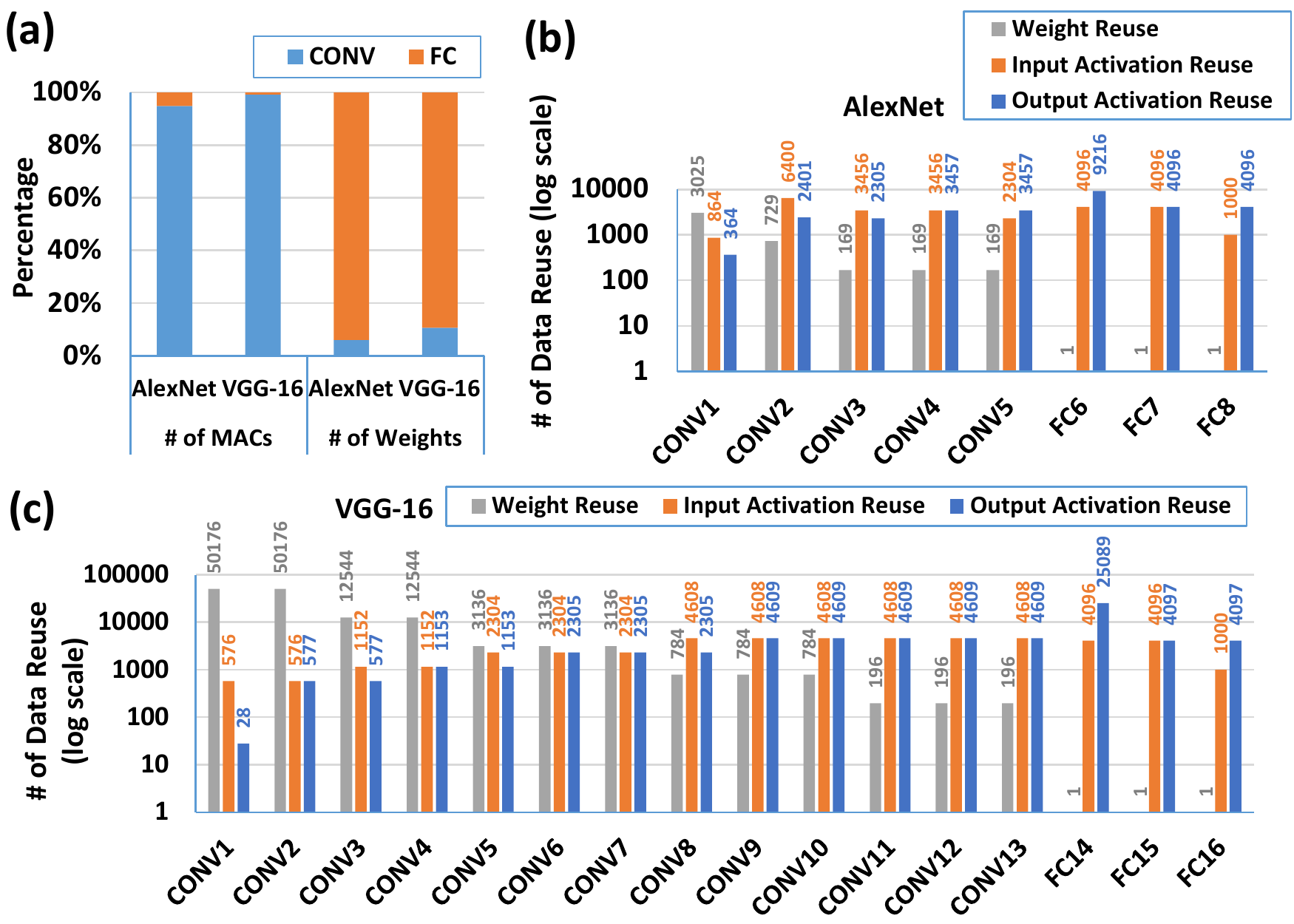}
\caption{Computing and Data-reuse statistics of the AlexNet and VGG-16.}
\label{fig:Fig_AlexNet_VGG_DataReuse}
\end{figure}

Table~\ref{table:DataStatistics} shows the number of MAC operations required for each CONV and FC layer of the AlexNet and VGG-16 networks. 
Fig.~\ref{fig:Fig_AlexNet_VGG_DataReuse}b and c show the weights-, input activations-, and output activations-reuse factor for the AlexNet and VGG-16, respectively. 
The {\it data-reuse factor} defines the number of MAC operations in which a specific data is used. It can be observed form the figures that the data reuse pattern can mainly be classified into two main categories: (1) CONV layers, where all types of data has significant reuse factor, and (2) FC layers, where per sample weight-reuse is 1. This classification is also supported by the fact that the CONV layers are more computationally intensive and the FC layers are more memory intensive, as shown  in Fig. ~\ref{fig:Fig_AlexNet_VGG_DataReuse}a. These observations along with the data-reuse pattern is  exploited in section~\ref{sec:MPNA}~and~\ref{sec:Dataflow} for designing a novel architecture that can maximally benefit from the data-reuse and a dataflow to reduce the number of off-chip memory accesses, respectively.

\begin{table}[htbp]
\caption{Number of MACs per input sample and weights in the AlexNet and VGG-16.}
\centering
\scriptsize
\begin{tabular}{|c|c|c|c|c|} 
\hline
\multirow{2}{*}{Observation} & \multicolumn{2}{|c|}{\# of MACs/Sample} & \multicolumn{2}{|c|}{\# of Weights}\\
\cline{2-5}
 & AlexNet & VGG-16 & AlexNet & VGG-16\\
\hline
\hline
CONV & 1.07B & 15.34B & 3.74M & 14.71M \\
\hline
FC & 58.62M & 123.63M & 58.63M & 123.64M\\ 
\hline
\end{tabular}
\label{table:DataStatistics}
\end{table}

%%%%%%%%%%%%%%%%%%%%%%%%%%%%%%%%%%%%%%%%%%%%%%%%%%%%%%%%%%%%%%%%%%%%%%%%%%%%%%%%%%%%%%%%%%%%%%%%%%%%%%%%%%%%%%%%%%%%%%%

\section{The MPNA Architecture}
\label{sec:MPNA}

\begin{figure*}[hbtp]
\centering
\includegraphics[width=1\linewidth]{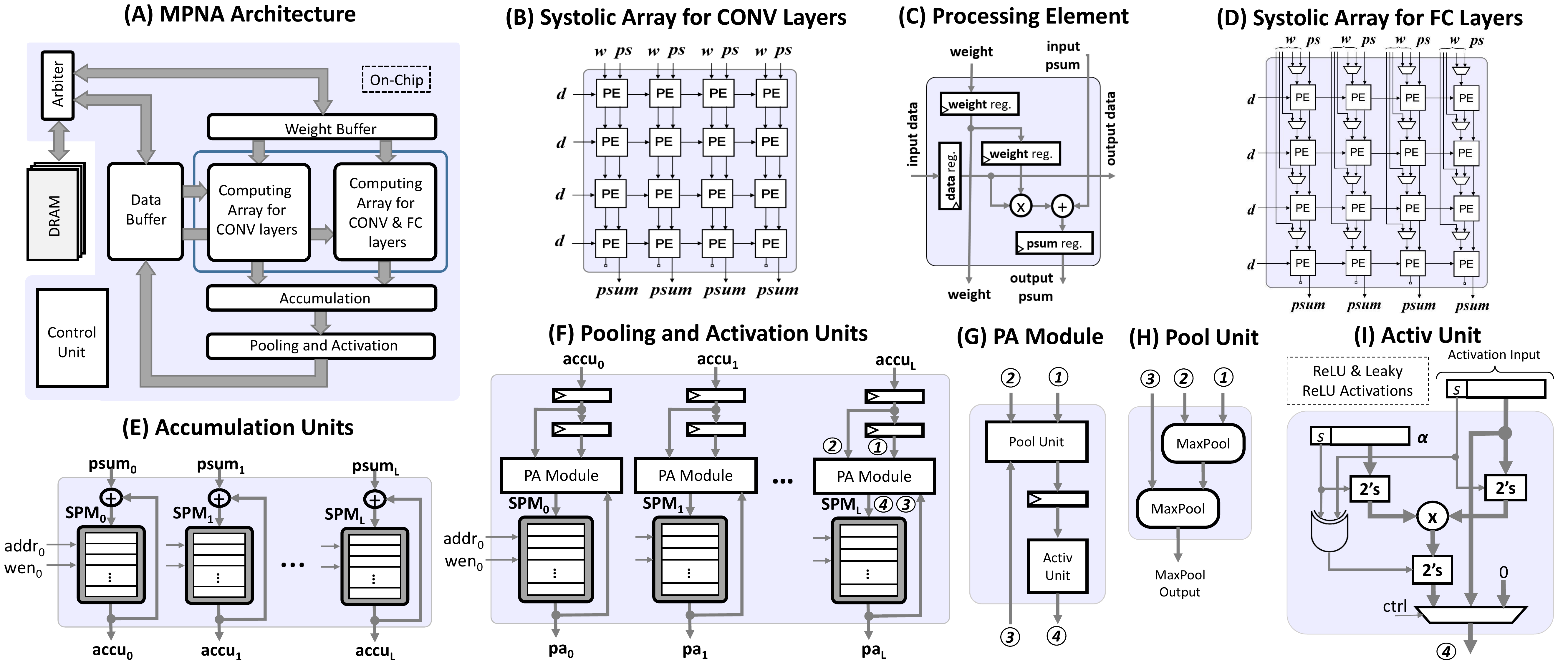}
\caption{MPNA architecture. (A) Top-level view of the architecture illustrating different components and their interconnection. Processing arrays (B) SA-CONV and (D) SA-FC for accelerating the convolutional and fully-connected layers, respectively. (C) Micro-architecture of a processing element (PE) of the systolic arrays. (E) Accumulation unit for storing and accumulating the partial sums generated by the systolic arrays. (F) Pooling and Activation unit for applying (H) MaxPooling and (I) ReLU/Leaky-ReLU activation function on the accumulated outputs.}
\label{fig:Fig_MPNA_Arch_Full}
\end{figure*}

Fig.~\ref{fig:Fig_MPNA_Arch_Full} presents the top-level view of our MPNA architecture with detailed components, as discussed in the subsequent sub-sections.

\subsection{Overview of our MPNA Architecture (Fig.~\ref{fig:Fig_MPNA_Arch_Full}A)}
The MPNA architecture is composed of two heterogeneous systolic arrays, an accumulation unit, a pooling \& activation unit, on-chip data and weight buffers, a control unit, and connectivity to DRAM. 
Each systolic array is specialized to support specific types of data parallelism for accelerating a specific set of configurations of computational layers while incurring minimum overheads. 
The systolic arrays receive data and weights from on-chip buffers, perform MAC operation, and forward the resultant partial sums to the accumulation block. 
The accumulation block is meant to hold the partial outputs while rest of their corresponding partial outputs are being computed, which are also then accumulated together inside the accumulator block. 
Once the output is complete, the accumulator block forwards the output(s) to the subsequent block for pooling and activation operation, or sends it back to the on-chip data buffer. 
The data is then either used for the next layer or is moved to the DRAM until rest of the intermediate operations are completed. 

\begin{figure*}[hbtp]
\centering
\includegraphics[width=1\linewidth]{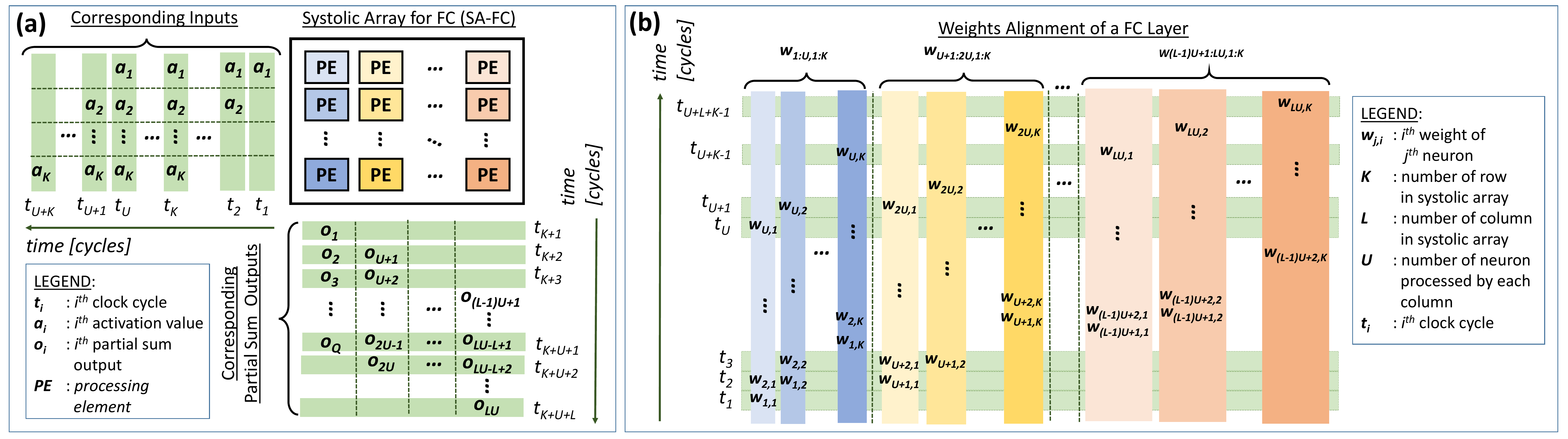}
\caption{Optimized Dataflow for FC layers: (a) Flow of input and output activations in SA-FC, (b) Alignment of weights fed into PEs (colored).}
\label{fig:Fig_MPNA_Dataflow_FC_Alignment}
\end{figure*}

\subsection{Heterogeneous Systolic Arrays (Fig.~\ref{fig:Fig_MPNA_Arch_Full}B-D)}

Based on the observations of Fig.~\ref{fig:Fig_AlexNet_VGG_DataReuse} in Section~\ref{sec:Analysis}, we conclude that there is a need for two heterogeneous systolic arrays for processing  different types of layers in a given CNN.

\textbf{Systolic Array for the Convolutional Layers (SA-CONV, Fig.~\ref{fig:Fig_MPNA_Arch_Full}B-C):} 
Following the advanced architectural trends in DNN systolic arrays like~\cite{Jouppi_GoogleTPU_ISCA17}, we design the CONV systolic array that can also exploit the activation (input data), weight, and partial sum reuse.
Our {\it SA-CONV} integrates a massively parallel array of Processing Elements (PEs) for dense MAC processing. Each PE receives the activations (input data) from its neighboring-left PE, and weights and partial sum from the neighboring-top PE, and passes the output to its downstream neighbor. 
The left-most PEs in the array receive data from the input buffers and the top-most PEs receive weights from the weight buffer. 
The processed data is then forwarded to the accumulation block by the bottom-most PEs. 
Such a systolic array enforces to have weights from the same filter/neuron to be mapped on the same column of the array, while the weights that are to be multiplied with the same input activations to be mapped on parallel columns. 
This enables high activation and weight reuse.

\textit{To support parallel weight movement during computation}, we proposed to include an additional register that can hold the weight values while the values which are to be used in the next iteration can be moved to their respective locations. This significantly reduces the initialization time of the systolic array. 

\textbf{Systolic Array for the Fully-Connected Layers (SA-FC, Fig.~\ref{fig:Fig_MPNA_Arch_Full}C-D):} 
As also supported by the studies of~\cite{Jouppi_GoogleTPU_ISCA17} for convolutional accelerators, the SA-CONV can provide significant throughput for multi-batch processing (with larger batch sizes), provided a reasonable size of on-chip memory.
However, this can significantly affect the latency of DNN inference which is an important parameter for almost all the real-world applications. To support such scenarios, we propose a novel systolic array architecture (SA-FC), which can accelerate both the CONV and the FC layers for {\it smaller batch sizes} as well. 
The design is based on the observation that the weight reuse factor-per-sample in all the FC layers is 1, as shown in Fig.~\ref{fig:Fig_AlexNet_VGG_DataReuse}b and c. This makes the SA-CONV ineffective as highlighted in Section~\ref{sec:Introduction}. However, the overall bandwidth required for such cases is huge, especially for larger DNNs. 
Therefore, our proposed systolic array {\it SA-FC} can be time-multiplexed for processing bandwidth intensive FC and computational intensive CONV layers. Towards generalization, it can also be effectively used for multi-batch processing while incurring minimum area and power overheads when compared to SA-CONV.
However, integrating both SA-FA and SA-CONV is a better design option w.r.t. the area, performance,and power/energy efficiency as we will show in the results section.
Fig.~\ref{fig:Fig_MPNA_Arch_Full}D shows that, unlike in SA-CONV, the SA-FC has dedicated connections from the weight buffer to each individual PE. This enables the system to update the weights in PEs at every clock cycle, and thereby providing the capability to support high-performance execution of the FC layers. The supporting dataflow for the SA-FC is illustrated in Fig.~\ref{fig:Fig_MPNA_Dataflow_FC_Alignment}.  

\subsection{Accumulation Unit (Fig.~\ref{fig:Fig_MPNA_Arch_Full}E)}

It is composed of several sub-units (equal to the total number of columns in SA-CONV and SA-FC) to support parallel processing. 
Each sub-unit is composed of a Scratch-Pad-Memory (SPM) for storing the partial output activations generated by the systolic arrays, and an adder for the accumulation of incoming partial sums with the stored values.
Once the output activations are complete, the values are forwarded to the succeeding pooling and activation block for further processing. 

\subsection{Pooling and Activation Unit (Fig.~\ref{fig:Fig_MPNA_Arch_Full}F-I)}

After the CONV and FC layers, the activation function is employed followed by a pooling layer that reduces the size of the feature maps for subsequent layers.
The MPNA provides support for the state-of-the-art {\it MaxPooling}, which is deployed in almost all the modern DNNs. 
Since the activation functions are typically monotonically increasing functions, they can be moved after the pooling operation to curtail the number activation functions to be performed and to reduce the hardware complexity. 
Figs.~\ref{fig:Fig_MPNA_Arch_Full}F-H show that this block consists of an SPM to hold the intermediate pooling results, and a pooling and activation computation module. 
The MPNA architecture currently supports two of best activation functions which are commonly used in DNNs (i.e., ReLU and Leaky-ReLU~\cite{Redmon_Yolo_CVPR16}).

%%%%%%%%%%%%%%%%%%%%%%%%%%%%%%%%%%%%%%%%%%%%%%%%%%%%%%%%%%%%%%%%%%%%%%%%%%%%%%%%%%%%%%%%%%%%%%%%%%%%%%%%%%%

\section{Dataflow Optimization}
\label{sec:Dataflow}
To effectively use the memory (both on-chip and off-chip) and the compute capabilities of our architecture, we propose a set of dataflows (Fig.~\ref{fig:Fig_MPNA_Dataflow_Tiling}) that can be employed depending on the configuration of the CONV and FC layers. To explain this, we first present the types of data-reuse and their dependencies on different data.

\subsection{Data-reuse and Their Dependencies}
\begin{itemize}
\item \textbf{Input Activation-Reuse} is the number of times an input activation is used by the same filter multiplied by the number of filters in a layer. To fully exploit this reuse, all the weights (i.e., $F^{i,:}$ where $i$ represents the feature map index of the input activation) and the corresponding output activations should be available for each available input sample.
\item \textbf{Output Activation-Reuse} is defined by the number of times partial sums are added into an output activation which, are determined by the size of the filters in a layer. To fully exploit this reuse, all the input activations and weight values corresponding to the output activation should be available.
\item \textbf{Weight-Reuse} is given as the number of times a weight value is used in the computation of a layer which equals the size of $OF^j$. To exploit this completely, all the input activations and the corresponding $OF$ map should be available on chip.
\end{itemize}

\begin{figure}[t]
\centering
\includegraphics[width=\linewidth]{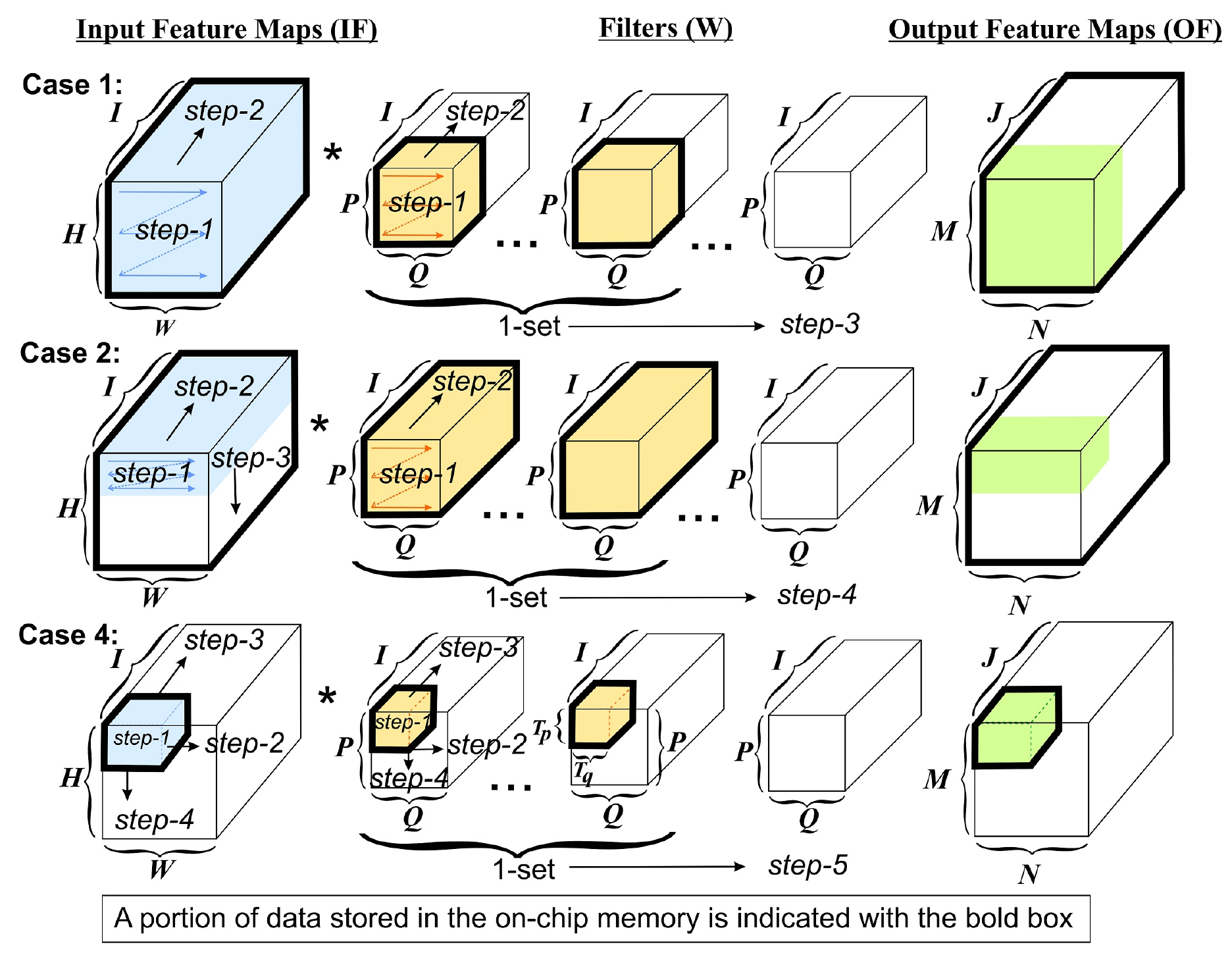}
\caption{Possible scenarios of dataflow patterns.}
\label{fig:Fig_MPNA_Dataflow_Tiling}
\end{figure}

\subsection{Possible Scenarios and Corresponding Dataflows (Fig.~\ref{fig:Fig_MPNA_Dataflow_Tiling})}
\textbf{Case 1:} All input and output activations, and a set of weights ($K \times L$) that has to be uploaded in the processing array in the next stage can be stored on-chip. Here, $K$ and $L$ represents the number of rows and columns in the systolic array. Also, the output activations in one $OF$ can be accommodated in the SPM of a single accumulation sub-unit. In this case, we can avoid input and output activation movement to DRAM and will fetch the weights once only. This is very effective for the later CONV layers where the total size of the input and output activation maps is small, and the number of filter parameters is huge.

\textbf{Case 2:} All input and output activations can be completely stored in on-chip data buffer. The output activations in one $OF$ cannot be accommodated in the SPM of a single accumulation sub-unit. In this case, if the overall weight buffer allows to accommodate $L$ (or $2$$\times$$L$) complete filters, we partition the input feature maps into multiple blocks to fit the output channels in SPMs, as shown in Fig.~\ref{fig:Fig_MPNA_Dataflow_Tiling}

\textbf{Case 3:} The Input and output activations cannot be completely stored on-chip. In this case, we give preference to input activations if they can be completely stored, and hence the Case 1 can be used while moving the resultant outputs to off-chip memory. 

\textbf{Case 4:} For all other cases, the best-possible configuration for partitioning data is selected using the methodology proposed in~\cite{Li_SmartShuttle} with following constraints: (1) The set of filters being processed together should be a multiple of $L$. (2) The number of weights selected from each filter at one time should be a multiple of $K$.

\subsection{Observations and Hardware Configurations}
We analyzed the configuration of the AlexNet~\cite{Krizhevsky_AlexNet_NIPS12} and defined our hardware configuration (shown in Table~\ref{table:Configurations}) on the following observations.
\begin{itemize}
\item The $OF$ of CONV$3$ till CONV5 (i.e., last three CONV layers) should fit in SPM of the accumulation, and the pooling and activation units. Since the size of $OFs$ in these layers is $13$$\times$$13$, we selected SPM which can hold up to $256$ elements.

\item For holding the input \& output activations of CONV$3$ till CONV$5$ layers of the AlexNet on-chip, we selected a $256$KB data buffer for two systolic arrays, i.e., greater than four times $13$$\times$$13$$\times$$384$ which is the size of the activation maps of CONV$4$.
\item Systolic array of size $8$$\times$$8$, which provides significant parallelism while not requiring much off-chip memory bandwidth.
\end{itemize}

\begin{table}[htbp]
\caption{MPNA Hardware Configurations.}
\centering
\scriptsize
\begin{tabular}{|c|c|} 
\hline
Module & Description \\
\hline
\hline
\multirow{2}{*}{Systolic Arrays} & Size of SA-CONV = 8x8 of PEs\\
 & Size of SA-FC = 8x8 of PEs\\
\hline
\multirow{2}{*}{SPM} & Size of SPM in each sub-unit of Accumulation block\\ 
& and Pooling \& Activation block = 256B\\
\hline
Weight Buffer & Size of weight buffer = 36KB\\
\hline
Data Buffer & Size of weight buffer = 256KB\\
\hline
\multirow{2}{*}{DRAM} & Size of DRAM = 2Gb\\
& Bandwidth of DRAM = 12.8GB/s \cite{Malladi_DRAM_ISCA12}\\
\hline
\end{tabular}
\label{table:Configurations}
\end{table}

%%%%%%%%%%%%%%%%%%%%%%%%%%%%%%%%%%%%%%%%%%%%%%%%%%%%%%%%%%%%%%%%%%%%%%%%%%%%%%%%%%%%%%%%%%%%%%%%%%%%%%%%%%%%%

\begin{figure}[t]
\centering
\includegraphics[width=\linewidth]{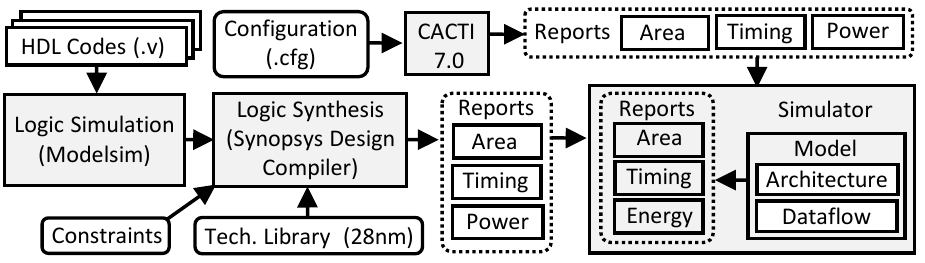}
\caption{Our tool flow and experimental methodology.}
\label{fig:MPNA_Flow}
\end{figure}

%%%%%%%%%%%%%%%%%%%%%%%%%%%%%%%%%%%%%%%%%%%%%%%%%%%%%%%%%%%%%%%%%%%%%%%%%%%%%%%%%%%%%%%%%%%%%%%%%%%%%%%%%%%%%%
\section{Evaluation Methodology (Fig.~\ref{fig:MPNA_Flow})}
\label{sec:Evaluation}
We developed a fully functional simulator to model the behavior of the MPNA. It is integrated with CACTI 7.0~\cite{CACTI7} for memory models and respective area, power,and energy estimation. 
The complete MPNA architecture is also designed in RTL and synthesized for a $28$nm technology using Synopsys design tools. We used ModelSim for logic simulation for functional and timing validations, and obtained the critical path delay, area, and power afterwards.
We compared our SA-FC architecture with SA-CONV, and our MPNA with conventional systolic array based accelerators (as baselines), for size of $2$$\times$$2$, $4$$\times$$4$, and $8$$\times$$8$ using AlexNet. 
We also compared our MPNA with several state-of-the-art accelerators such as Eyeriss~\cite{Chen_Eyeriss_ISSCC16}, SCNN~\cite{Parashar_SCNN_ISCA17}, and FlexFlow~\cite{Lu_FlexFlow_HPCA17}.

\begin{figure}[t]
\centering
\includegraphics[width=1\linewidth]{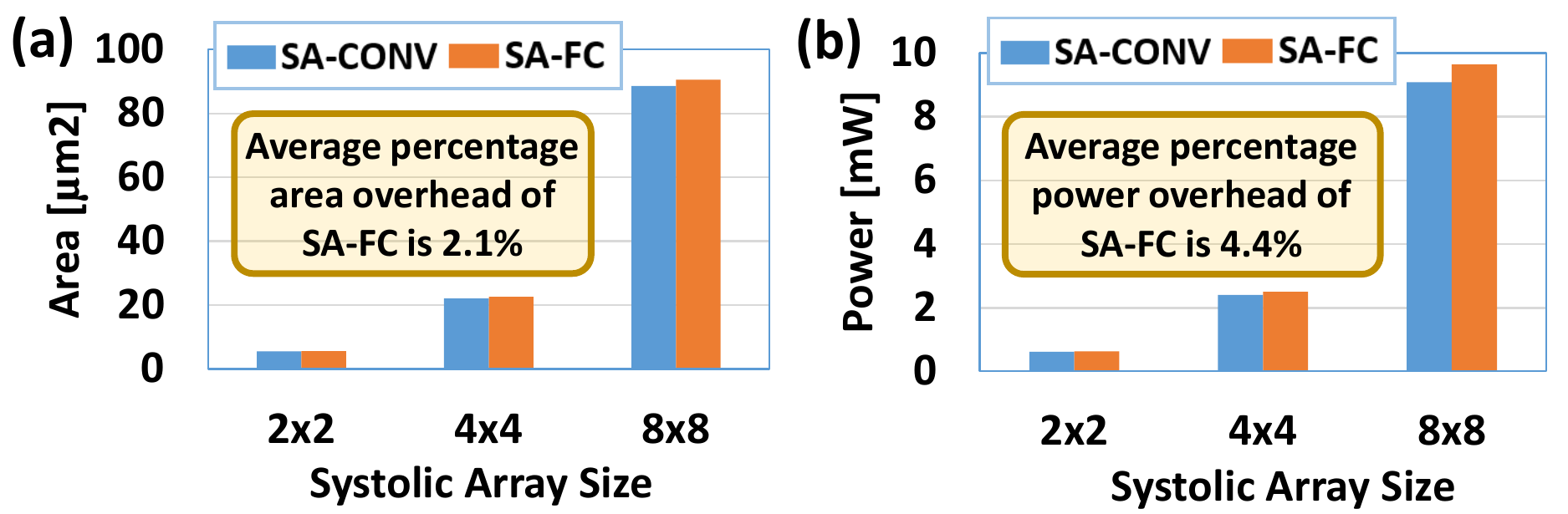}
\caption{Comparing the conventional systolic array and the proposed SA-FC in terms of (a) area and (b) power.}
\label{fig:Fig_SAvsISA_AreaPower}
\end{figure}

%%%%%%%%%%%%%%%%%%%%%%%%%%%%%%%%%%%%%%%%%%%%%%%%%%%%%%%%%%%%%%%%%%%%%%%%%%%%%%%%%%%%%%%%%%%%%%%%%%%%%%%%%%%%%%
\section{Experimental Results and Discussion}
\label{sec:Results}

\begin{figure*}[t]
\centering
\includegraphics[width=1\linewidth]{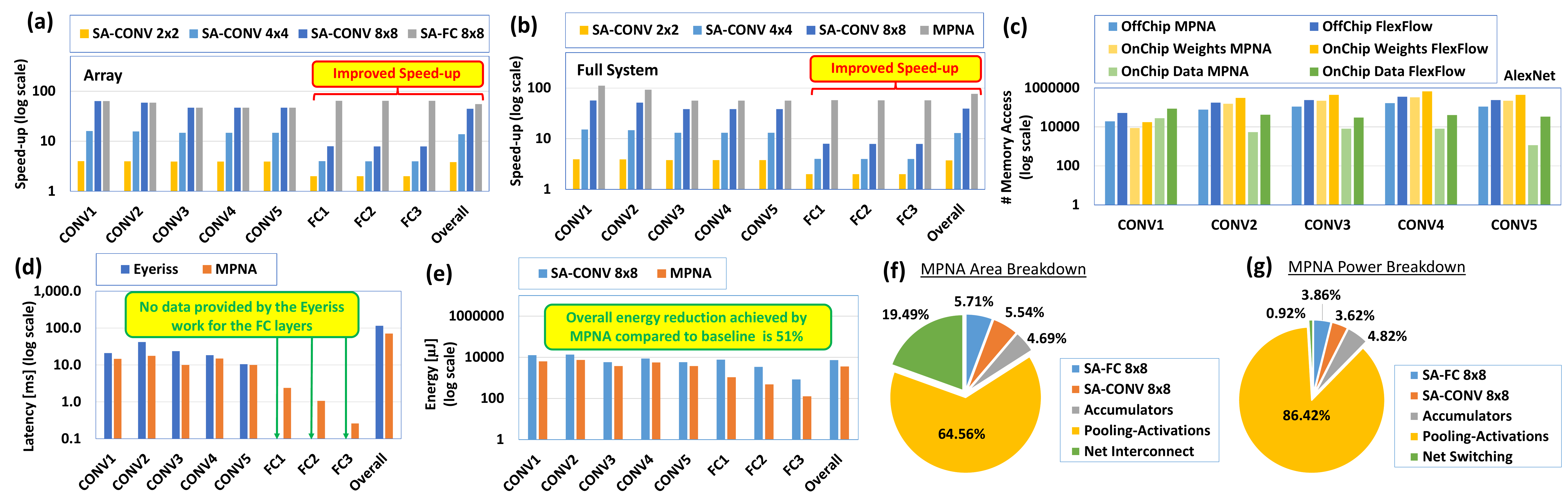}
\caption{Evaluation using AlexNet on (a) speedup of different types of systolic arrays; (b) speedup of MPNA vs conventional architectures with different systolic arrays; (c) estimated number of off-/on-chip memory accesses of MPNA vs. FlexFlow~\cite{Lu_FlexFlow_HPCA17}; (d) estimated processing latency of MPNA vs. Eyeriss~\cite{Chen_Eyeriss_JSSC16}; and (e) energy-efficiency of MPNA vs baseline architecture. (f-g) Area and power breakdown of MPNA hardware.}
\label{fig:Fig_Results_AlexNet}
\end{figure*}

%%%%%%%%%%%
\textbf{Systolic Array:} We compared our SA-FC and SA-CONV to get the profile of the proposed SA-FC in terms of area and power. 
%The area results presented in 
Fig.~\ref{fig:Fig_SAvsISA_AreaPower} shows that the SA-FC incurs insignificant area and power overhead ($2.1\%$ and $4.4\%$, respectively) compared to SA-CONV.
Fig.~\ref{fig:Fig_Results_AlexNet}a shows that SA-FC achieves 8.1$\times$ speed-up, compared to when only using the SA-CONV for FC layers, due to its microarchitural enhancements that can provide the data timely to PEs for generating results each clock cycle.

\textbf{Key Observations for Performance Evaluation (Figs.~\ref{fig:Fig_Results_AlexNet}b-d):} 
\begin{itemize}
\item Our MPNA achieves 1.4$\times$ -- 7.2$\times$ higher speed-up for AlexNet compared to the conventional systolic array-based architectures. These improvements come from the parallelism of heterogeneous computing arrays and efficient dataflows.

\item Compared to FlexFlow~\cite{Lu_FlexFlow_HPCA17}, our MPNA requires 53\% less number of memory accesses (Fig.~\ref{fig:Fig_Results_AlexNet}c) due to our optimized dataflows, leading to significant performance and energy improvements.

\item Compared to Eyeriss~\cite{Chen_Eyeriss_JSSC16}, our MPNA achieves 1.7$\times$ better latency for CONV layers, as well as significant speedup for FC layers (see Fig.~\ref{fig:Fig_Results_AlexNet}d), while Eyeriss does not disclose their latency results for the FC layers.

\item Further comparison of MPNA with the state-of-the-art accelerators is summarized in Table~\ref{table:CompareSoA}, showing competitive characteristics of our MPNA for full CNN acceleration.
\end{itemize}

%%%%%%%%%
\textbf{Key Observations for Power/Energy Evaluation (Figs.~\ref{fig:Fig_Results_AlexNet}e,g):} 
\begin{itemize}
\item MPNA consumes $239$mW average power, which is dominated by pooling and activation unit due to its local memories and activation function processing.

\item MPNA achieves overall 51\% of energy reduction compared to baseline architecture (Fig.~\ref{fig:Fig_Results_AlexNet}e) due to reduced memory access and maximal data-reuse as a result of optimized dataflows.
\end{itemize}

%%%%%%%%%
\textbf{Key Observations for Area Evaluation (Figs.~\ref{fig:Fig_Results_AlexNet}f):} 
The area of MPNA ($2.34$mm$^2$) is occupied by computational parts ($1.38$mm$^2$) and on-chip memories ($0.96$mm$^2$) comprising data and weights buffers. 
Table~\ref{table:CompareSoA} shows that our MPNA consumes a competitively small area compared to other state-of-the-art accelerators. 

\begin{table}[t]
\caption{Comparison of the state-of-the-art accelerators.}
\centering
\scriptsize
\begin{tabular}{|c|c|c|c|c|} 
\hline
\multirow{2}{*}{Reference} & Eyeriss & SCNN & FlexFlow & MPNA\\
 & \cite{Chen_Eyeriss_JSSC16} & \cite{Parashar_SCNN_ISCA17} & \cite{Lu_FlexFlow_HPCA17} & (this work)\\
\hline
\hline
Technology (nm) & 65 & 16 & 65 & 28 \\ 
\hline
Precision (fixed-point) & 16-bit & 16-bit & 16-bit & 8-bit\\ 
\hline
\# PEs & 168 & 64 & 256 & 128 \\
\hline
On-chip Memory (KB) & 181.5 & 1024 & 64 & 288\\
\hline
Area (mm$^2$) & 12.25 & 7.9 & 3.89 & 2.34\\
\hline
Power (mW) & 278 & NA & $\sim$1000 & 239\\
\hline
Frequency (MHz) & 100-250 & 1000 & 1000 & 280\\
\hline
Performance (GOPS) & 23.1 & NA & 420 & 35.8\\
\hline
Efficiency (GOPS/W) & 83.1 & NA & 300-500 & 149.7\\ 
\hline
Acceleration Target & CONV & CONV & CONV & CONV+FC\\
\hline
\end{tabular}
\label{table:CompareSoA}
\end{table}

%%%%%%%%%%%%%%%%%%%%%%%%%%%%%%%%%%%%%%%%%%%%%%%%%%%%%%%%%%%%%%%%%%%%%%%%%%%%%%%%%%%%%%%%%%%%%%%%%%%%%%%%%%%%%%
\section{Conclusion}
\label{sec:Conclusion}
In this work, we demonstrate that a significant speedup for \textit{both} CONV and FC layers can be achieved by a synergistic design methodology encompassing dataflow optimization, diverse types of data-reuse and the MPNA architecture with heterogeneous systolic arrays and specialized buffers. The complete architecture is synthesized in a $28$nm technology with ASIC design flow, and a comprehensive evaluation is done for area, performance, power, and energy, showing significant gains of our approach over various state-of-the-art. Our novel concepts and open-source hardware would enable further research on accelerating emerging DNNs (like Capsule neural networks).

%%%%%%%%%%
%%%%%%%%%%%%%%%%%%%%%%%%%%%%%%%%%%%%

\end{document}